\documentclass[12pt]{iopart}

\usepackage{graphicx}
\usepackage{iopams}

\begin{document}
\bibliographystyle{prsty}

\title{Kinetic roughening in a realistic model of non-conserved interface growth}

\author{Matteo Nicoli}
\address{Dpto.\ Matem\'aticas and Grupo Interdisciplinar de Sistemas Complejos (GISC), Universidad Carlos III de Madrid, Avenida Universidad 30, 28911 Legan\'{e}s, Spain}

\author{Mario Castro}
\address{Escuela T\'ecnica Superior de Ingenier\'{\i}a (ICAI) and GISC \\
Universidad Pontificia Comillas, E-28015 Madrid, Spain}

\author{Rodolfo Cuerno}
\address{Dpto.\ Matem\'aticas and GISC, Universidad
Carlos III de Madrid, Avenida Universidad 30, 28911 Legan\'{e}s, Spain}

\begin{abstract}
We provide a quantitative picture of non-conserved interface growth from a diffusive field making special emphasis on two main issues, the range of validity of the effective small-slopes (interfacial) theories and the interplay between the emergence of morphologically instabilities in the aggregate dynamics, and its kinetic roughening properties. Taking for definiteness electrochemical deposition as our experimental field of reference, our theoretical approach makes use of two complementary approaches: interfacial effective equations and a phase-field formulation of the electrodeposition process. Both descriptions allow us to establish a close quantitative connection between theory and experiments.
Moreover, we are able to correlate the anomalous scaling properties seen in some experiments
with the failure of the small slope approximation, and to assess the effective re-emergence of standard kinetic roughening properties at very long times under appropriate experimental conditions.

\end{abstract}


\maketitle 

\section{Introduction}

Many natural and technological growth systems exist whose dynamics results
from the interplay between diffusive transport of aggregating units within
dilute phase, and their attachment at a finite rate to a condensed growing
cluster. Examples abound particularly within Materials Science, e.g.\ in the
production of both amorphous (through Chemical Vapor Deposition \cite{cvd}
or Electrochemical Deposition \cite{ecd}), and of epitaxial (by Molecular
Beam Epitaxy \cite{mbe}) thin films. Still, a similar variety of surface
morphologies can be found in diverse systems, such as e.g.\ bacterial colonies
\cite{matsushita2005cfb}, to the extent that universal mechanisms are expected
to govern the evolution of these non-equilibrium systems \cite{ball}.
Moreover, many times it is not matter, but e.g.\ heat that is transported, but
obeying precisely the same constitutive laws (one-sided solidification).

The objects of interest in the time evolution of these systems are the
morphological properties of the aggregate which is produced. Thus, a variety
of behaviors ensue, such as pattern formation and evolution \cite{ball},
and surface kinetic roughening \cite{barabasi}, whose understanding has driven
a large effort within the Statistical Mechanics and Nonlinear Science
communities. In this context, continuum models have played a prominent
role, due to the compact description they typically provide of otherwise
very complex phenomena, and their ability to explore generic or universal
system properties \cite{cuerno:2007}. For the diffusive growth systems
mentioned above, the simplest continuum formulation takes the form of a moving
boundary problem in which particle diffusion in the dilute phase is coupled
with a boundary condition in which the growth velocity of the aggregate is
prescribed as, roughly, the fraction of random walkers that finally stick
irreversibly to it. Thus, the evolution of a growing surface from a vapor
or the electrochemical growth of a metallic deposit under galvanostatic
conditions can be described by the same nonlinear system of stochastic differential
equations (see Sec.\ \ref{cm} for details).

Electrochemical deposition (ECD) has been widely studied mainly due to
the apparent simplicity of the experimental setup and the capability to
control different relevant physical parameters. Most of the studies in
ECD address the growth of quasi-two dimensional fractal aggregates. This
can be achieved by confining a metallic salt between two parallel plates
with a small separation between them. Thus, although the formulation of the problem can be accurately posed in order to include all the physical mechanisms involved, its solution is far from trivial (actually it cannot be obtained even in the deterministic
case). Moreover, the numerical integration of the corresponding equations often involves the
use of complex algorithms that take into account the motion of the boundary
and which, in practice, are difficult to implement. Besides this intrinsic
complexity, the role of fluctuations and their inclusion in the numerical
scheme is still an open question. In the present context, the interplay
among different transport or electrochemical mechanisms is not clear and the
predictive power of different models is limited to the description of the
concentration of cations in the solution or the qualitative morphologies of
the fractal aggregates~\cite{castro1998asn,castro2000mbd}.

In this scenario, two possibilities arise from a theoretical point of view. On the one hand,
one can try to obtain as much information as available from customary
perturbation methods. Thus, linearization, Green's function projection
techniques or weakly nonlinear expansions provide important information
at both short (linear regime) and asymptotic times~\cite{nicoli:2008}.
Briefly, this approach allows to explore the physics of the problem in
the so-called small slopes approximation. On the other hand, one can try to
integrate numerically a more computationally efficient version
of the moving boundary problem in order to accurately determine the validity
of the approximations in the model formulation, and also to gain insight into the intermediate time regimes.

This second approach can be achieved by the introduction of an equivalent
phase-field model~\cite{gonzalez-cinca:2004} version of the moving
boundary problem. Phase-field models have been successfully introduced
in the last few years to understand moving boundary problems
which are harder to integrate numerically (or in combination with discrete
lattice algorithms~\cite{plapp2000mrw}). Moreover, they have become a
paradigmatic representation of multi-phase dynamics whose application
ranges from solidification~\cite{karma1996pfm,wang1993tcp} or Materials
Science~\cite{elder2002mec,granasy2003gdd,karma1998ssg,castro2003pfa}
to Fluid Mechanics~\cite{dube2000cda,alanissila2004pfm} or even
Biology~\cite{campelo2007mcd}.


The aim of this paper is to provide a quantitative picture of non-conserved growth in which material transport is of a diffusive nature. Working for the case of ECD, we will focus on two important and complementary tasks: assessing the validity of the effective small-slopes
(interfacial) theories, and the emergence of unstable aggregates that cannot be described by single valued functions. More importantly, we will show that our model is capable to provide both a qualitative and a quantitative picture of growth by electrochemical deposition.
In particular, our present study allows us to provide more specific reasons for the experimental difficulty in observing kinetic roughening properties of the Kardar-Parisi-Zhang universality class, in spite of theoretical expectations. Finally, we will see that the
frequently observed anomalous scaling properties arise in this class of systems associated with the failure of the single-valued description of the interface, and may yield asymptotically to standard scaling behavior, provided a well defined aggregate front reappears.

Our paper is organized according to the following scheme: in Sec.\ II we
introduce our model and discuss in depth the relationship between its defining
parameters and physical ones. In Sec.\ III, we discuss the validity of the
small slope approximation and conclude on the need of a full phase-field
description of the problem. Section IV collects the results of a detailed comparison with specific ECD experiments that allows us to derive a consistent picture of pattern formation within our non-conserved growth system. Finally, Sec.\ IV concludes the
paper by summarizing our main findings and collecting some prospects and open
questions.

\section{Continuum model of galvanostatic electrodeposition}
\label{cm}

Recently we have proposed a stochastic moving boundary model to describe
the morphological evolution of a variety of so-called diffusive growth
systems \cite{nicoli:2008}. This is a large class of systems in which
dynamics arises as a competition between supply of material from a dilute
phase through diffusion, and aggregation at a growing cluster surface
through generically finite attachment kinetics. The deterministic limit of
the model has a long history in the field of thin film growth by Chemical
Vapor Deposition \cite{Jansen,Palmer88,Bales}, where diffusing species are
electrically neutral, conceptually simplifying the theoretical description.
Naturally, in the case of aggregate growth by electrochemical deposition
(ECD) in galvanostatic (constant current) conditions, the situation is
more complex due to the existence of two diffusing species (anions and
cations) and an imposed electric field. Nonetheless, an explicit mapping
of the corresponding dynamical equations can be made to the CVD system,
by means of the following assumptions: (i) zero fluid velocity in a thin
electrochemical cell; (ii) anion annihilation at the anode, whose position is
located at infinity, namely, the aggregate height is much smaller than the
distance separating the electrodes; (iii) electroneutrality away from the
electrodes; (iv) zero anion flux at the cathode. Within these assumptions,
the model ECD system takes the form \cite{nicoli:2008}
\begin{eqnarray}
\partial_tc&=& D\nabla^2c-\nabla\cdot{\bf q},\label{n_difusion_eq}\\
D\partial_nc\Big|_\zeta &=&k_D(c-c^0_{eq}+\Gamma\kappa+\chi)\Big|_\zeta+
{\bf q}\cdot{\bf n},\label{n_boundary_cond}\\ V_n&=&\Omega \, \Big[
D\partial_nc-\nabla_s\cdot{\bf J}_s -{\bf q}\cdot{\bf n} -\nabla_s\cdot{\bf
p}\Big], \label{n_velocidad} \\ \lim_{z\rightarrow\infty} & &
\hskip -0.5 truecm c(x,z;t) =c_a.\label{C_anode}
\end{eqnarray} In
\eref{n_difusion_eq}-\eref{C_anode} the field $c$ is the cation concentration
$C$, rescaled by a quantity $R_c$ that depends on the diffusivity and
the mobility of the two species (cations and anions). Specifically,
$R_c$ is the ratio between the ambipolar diffusion coefficient $D$ and
the constant $t_c$, defined by \begin{equation} D=\frac{\mu_c D_a+\mu_a
D_c}{\mu_c+\mu_a}\qquad t_c=\frac{\mu_c}{\mu_c+\mu_a}, \end{equation}
where $\mu_a(\mu_c)$, $D_a(D_c)$ are the anion (cation) mobility and
diffusivity, respectively. Other parameters in Eqs.\ \eref{n_difusion_eq} to
\eref{C_anode} are $c^0_{eq}=R_c C_a \exp[z_c F\eta/(RT)]$, $c_a=R_c C_a$,
$\Gamma=\gamma/(RT)$, $\Omega=M/\rho=(1-t_c)/C_a$, and the mass transport
coefficient $k_D=J_0\exp[-bz_c F\eta/(RT)]/(z_c F C_a R_c)$, where $C_a$ is the
initial cation concentration, held fixed at infinity, $M$ is the metal molar
mass and $\rho$ the aggregate mean density, $T$ is temperature, $\gamma$
is the aggregate surface tension, $R$ is the ideal gas constant and $F$ is
Faraday's constant, $ez_c$ is the cation charge, $J_0$ is the exchange current
density in equilibrium, and $\eta$ is the overpotential from which a surface
curvature contribution in the corresponding Butler-Volmer boundary condition
\cite{ecd,Bard,Kashchiev} has been singled out, parameter $b$ estimating the asymmetry
of the energy barrier related to cation reduction.

Eq.\ \eref{n_difusion_eq} describes cation diffusion in the cell, with
a noise $\mathbf{q}$ term that describes fluctuations in the (conserved) diffusive
current. Eq.\ \eref{n_boundary_cond} is a mixed boundary
condition at the aggregate surface $\zeta$ that determines the efficiency
at which the cation flux leads to actual aggregation. It originates from
the standard Butler-Volmer condition, and allows for a noise term $\chi$
accounting for fluctuations in attachment. In this equation $\partial_n$
represent the derivative along the normal component $\bf{n}$ of the
interface. Eq.\ \eref{n_velocidad} computes
the growth velocity along the local normal direction $V_n = \mathbf{V} \cdot
\mathbf{n}$, incorporating the contributions from the diffusive current
mediated by the boundary condition \eref{n_boundary_cond}, as well as
from other diffusive currents at the surface, $\mathbf{J}_{s}$, in which
fluctuations are described by the noise term $\mathbf{p}$. The standard
choice is that of surface diffusion \cite{mullins:1957,mullins:1959},
$\nabla_s\cdot\mathbf{J}_{s}=\Omega^2\nu_s D_s\gamma(k_B T)^{-1}\nabla_s^4
\zeta$, where $\nu_s$ is the surface concentration of diffusing species and
$D_s$ is their surface diffusivity. Here, $\nabla_s^2$ denotes the surface
Laplacian. Finally, the stochastic moving boundary problem is complete with
the boundary condition \eref{C_anode} prescribing a fixed cation concentration
at the anode.

In the above model of ECD, in principle all parameters can be estimated in
experiments. This is also the case for the noise terms: indeed, their amplitudes
can be shown to be functions of the physical parameters above, once we take
the noises as Gaussian distributed numbers that are uncorrelated in time and
space, and make a local equilibrium approximation \cite{nicoli:2008}. The
mathematical problem posed by Eqs.\ \eref{n_difusion_eq} to \eref{C_anode}
is analytically very hard to work with. However, substantial progress
can be made by means of a small slope, approximation $|\nabla \zeta|\ll
1$ \cite{ourprl,cuerno:2002,nicoli:2008}. This approach allows to obtain
a closed evolution equation for the Fourier modes $\zeta_k(t)$ of
disturbances of the surface height away from a flat interface. Generically,
such an equation takes the form \cite{ourprl,cuerno:2002,nicoli:2008}
\begin{equation}
\partial_t\zeta_k(t)=\omega_k\zeta_{k}(t)+
\frac{V}{2}\mathcal{N}[\zeta]_k+\eta_k(t),
    \label{pre_KPZ}
\end{equation}
where $\eta_k(t)$ is an additive noise term including both conserved and
non-conserved contributions, and $\mathcal{N}[\zeta]_k$ is the Fourier
transform of $\mathcal{N}[\zeta]=(\partial_x\zeta)^2$, namely, of the celebrated
Kardar-Parisi-Zhang (KPZ) nonlinearity that is the landmark of non-conserved
interface growth. Here, $V$ is the mean velocity of a flat interface, while
the dispersion relation $\omega_k$ is a function of wave-vector magnitude $k$
that provides the (linear) rate of growth or decay of a periodic perturbation
$\zeta_k(t)$ of the flat solution. In the simplest long-wavelength limit,
the dispersion relation becomes a polynomial in $|k|$, with two important
limits, namely
\begin{equation}
\label{omega_poly} \omega_k = \left\{
\begin{array}{lc} V|k|[1-(d_0l_D+B\Omega/2D)k^2]-B\Omega k^4, & k_D \to
\infty \quad \mbox{MS} \\ (Dk_D\Delta/V)( k^2-l_D d_0 [1-(Vd_0/D)^{1/2}]^{-1}
k^4 ), & k_D < \infty \quad \mbox{KS} \end{array} \right.
\end{equation}
where $B=\Omega^2\gamma\nu_s D_s/(RT)$, and $\Delta=1-d_0/l_D$ with
$d_0=\Gamma \Omega$ the capillary length and $l_D=D/V$ the diffusion
length, the condition $d_0\ll l_D$ applying in most physical cases. As we
see from \eref{omega_poly}, in all cases the dispersion relation predicts
a morphological instability in which there is a band of linearly unstable
Fourier modes whose amplitudes grow exponentially in time. One of these
modes grows at a faster rate than all the other, providing the selection of a characteristic
length-scale in the surface and formation of a periodic pattern. However,
some features of the instability change as a function of the mass transfer
parameter $k_D$. Thus, when attachment kinetics is very fast ($k_D\to\infty$)
the dispersion relation has the classic shape of the Mullins-Sekerka (MS)
instability. In this case, Eq.\ \eref{pre_KPZ} is non-local in space, as
a reflection of the strong implicit shadowing effects associated with
diffusion. In the case of slow kinetics and, as long as $d_0 < l_D$, we still
have a morphological instability, but of a local nature, Eq.\ \eref{pre_KPZ}
becoming the celebrated (stochastic) Kuramoto-Sivashinsky (KS) equation in
real space.

Remarkably, the KPZ nonlinearity is able to tame the morphological
instabilities in \eref{pre_KPZ} and induce, albeit after relatively
long transients, kinetic roughening properties at long length scales
\cite{nicoli:2008}. These are those of the KPZ equation for slow attachment
kinetics, while they correspond to a different universality class for
$k_D\to\infty$. Thus, one would expect KPZ scaling for a wide range of
parameters, which has indeed been found experimentally within ECD growth \cite{Schilardi}, but not with the generality that was expected
\cite{cuerno:2002,cuerno:2004}. One possibility is that the
crossovers induced by the instabilities are so long lived as to prevent
actual asymptotic scaling from being accessed by experiments. Such a possibility
seems to occur in conceptually similar growth techniques, such as Chemical
Vapor Deposition \cite{Ojeda,ojeda:2003}. However, standard symmetry
arguments \cite{barabasi} lead to the expectation of generic KPZ behavior
for such non-conserved growth phenomena as provided by ECD systems. Thus,
in order to understand this disagreement, in the next section we undertake
a detailed comparison between the small slope approximation and ECD growth
experiments.

\section{Dispersion Relation and Small Slope Approximation}


The simplest way to improve upon the analytical results of the previous section is to realize that Equation \eref{omega_poly} merely provides a long-wavelength approximation of the exact linear dispersion relation for the moving boundary problem. We can consider the full shape of the dispersion relation, that turns out to be an implicit (rather than explicit) function of wave-vector, namely \cite{nicoli:2008b}
\begin{equation}
\left(1+\left[1+\hat{\omega}_k+\hat{k}^2\right]^{1/2}\right)
\left(2-\frac{V}{2k_D}\hat{\omega}_k-\frac{d_0}{2 l_D} \hat{k}^2\right)
    =\hat{\omega}_k+4,
\label{omega_compl}
\end{equation}
where $\hat{k}=2D k/V$ and $\hat{\omega}_k=4D\omega_k/V^2$. Using this formula, we can see the extent to which surface kinetics modifies the functional form of the linear dispersion relation. Actually, several experimental studies of ECD have already tried to estimate the functional form of the dispersion relation in compact aggregates.
An example taken from the literature is the growth of
quasi-two-dimensional Cu and Ag branches by de Bruyn~\cite{bruynpa,bruynpre}.
In \cite{bruynpre}, the branches grow in the cathode due to the
electrodeposition of ions created from the CuSO$_4$ aqueous solution.
The analysis of the early stages of the branch growth allows the
author to fit the experimental dispersion relation to
\begin{equation}
\omega_k=\frac{q|k|(1-rk^2)}{1+sk},
\label{omegabruyn}
\end{equation}
where $q$, $r$ and $s$
depend on the properties of the electrolyte and the deposited metal, the ion
concentration, and the applied current. From the reported data and figures in Ref.\ \cite{bruynpre} we can extract the information summarized in Table \ref{tablebruyn}, where we also include data from the experiments in \cite{kahanda,Leger2}.
\begin{table}[!h]
\begin{center}
\begin{tabular}{|c|c|c|c|} \hline
 {\bf Parameter } &{\bf Ref.\ \cite{bruynpre}} & {\bf  Ref.\ \cite{kahanda}} & {\bf Ref.\ \cite{Leger2}} \\ \hline
 $C_a$ [mol/cm$^3$] & $1 \times 10^{-4}$ &$2 \times 10^{-3}$ &$5 \times 10^{-4}$ \\ \hline
 $I$ [mA]&$1.4225$ &NR & NR\\ \hline
$J_0$ [mA/cm$^2$]&$3\times 10^{-2}$  & $3\times 10^{-2}$& $3\times 10^{-2}$\\ \hline
 $J$ [mA/cm$^2$] &$11.16$  & NR & 65.00 \\ \hline
 $D_c$ [cm$^2$/s]& $0.720\times 10^{-5}$  &$0.720\times 10^{-5}$ &$0.720\times 10^{-5}$ \\ \hline
 $D_a$ [cm$^2$/s]& $1.065\times 10^{-5}$  &$1.065\times 10^{-5}$ &$1.902\times 10^{-5}$ \\ \hline
 $z_c$&$2$ & 2&2 \\ \hline
 $V$ [cm/s] & $3.45 \times 10^{-4}$ &$\approx 1.00\times 10^{-6}$  &$4.87 \times 10^{-4}$ \\ \hline
 $l_D$ [cm]&$ 2.49\times 10^{-2}$ &$\simeq 8.59$ &$2.15\times 10^{-2} $\\ \hline 
\end{tabular}
\caption{Experimental parameter values estimated from Refs.\ \cite{bruynpre} (second column), \cite{kahanda} (third column), and \cite{Leger2} (4th.\ column). NR stands for not reported.}
\label{tablebruyn}
\end{center}
\end{table}
We need to supplement these values with reasonable estimates for the capillary
length, $d_0$ and the mass transfer coefficient, $k_D$. We estimate $d_0$ from
the experiments by Kahanda {\em et al.}~\cite{kahanda} (and summarized in
Table \ref{tablebruyn}; see also below) and we rescale this value for
the right concentration to $d_0=2\times 10^{-5}$ cm.
Finally, $k_D$ can be obtained from the location of the maximum in the
dispersion relation. Thus we can estimate $k_D$ to be in the range (due to the
experimental uncertainty) $4.44-5.56 \times 10^{-4}$ cm/s.

\vskip 0.5cm
\begin{figure}[!h]
\begin{center}
\includegraphics[width=.6\textwidth]{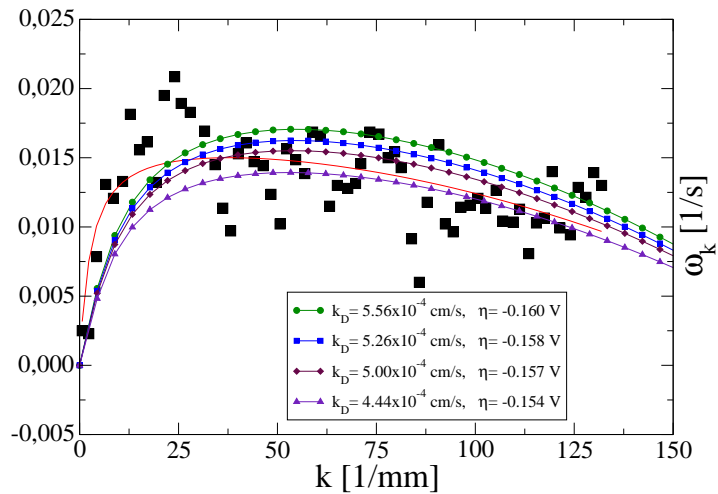}
\end{center}
\caption{Linear dispersion relation, $\omega_k$, of Cu electrodeposition. Squares provide the experimental values taken from Ref.\ \cite{bruynpre}, and the red line is the fit to  \eref{omegabruyn} made in this reference. The other lines are numerical solutions of  \eref{omega_compl} using the parameters in the second column in Table \ref{tablebruyn}
for different values of $k_D$ as indicated in the legend.}
\label{bruyn_pre_fig}
\end{figure}

These data are compatible with the assumptions made in the derivation of the
moving boundary problem introduced in Ref.\ \cite{nicoli:2008}. For instance,
the diffusion length is larger than the branch width (typically $0.1$ mm
in these experiments) but the mean growth velocity is comparable to the
mass transport coefficient, and surface kinetics become relevant.
For this reason, the growth process is not diffusion limited and the
dispersion relation is very different from the Mullins-Sekerka one. Similarly,
the estimate for $k_D$ implies that the applied overpotential $\eta$ varies
between $-0.160$ V and $-0.154$ V, which are plausible enough. With this
information we can now proceed to recast the experimental results in Ref.\
\cite{bruynpre} as shown in Fig.\ \ref{bruyn_pre_fig}.
The agreement between our theory \eref{omega_compl} and experiment is remarkable, taking into account that we are not performing any fit of the data. Moreover, the long wavelength approximations \eref{omega_poly} are not able to fit the experimental data.

\begin{figure}[!h]
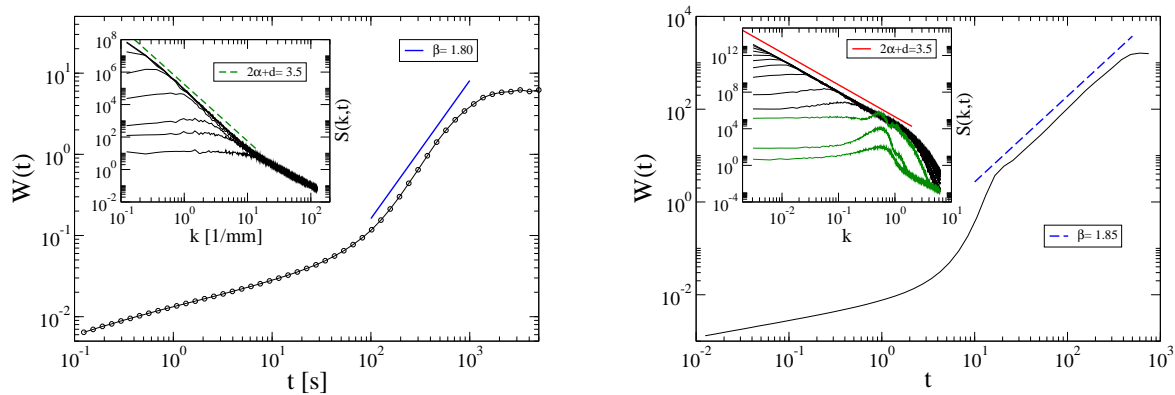

\begin{center}
\begin{tabular}{cc}
\begin{minipage}{0.5\textwidth}
\includegraphics[width=0.9\textwidth]{W.psd.deBruyn.eps}
\end{minipage}
&\begin{minipage}{0.5\textwidth}
\includegraphics[width=0.9\textwidth]{W_psd_mu075.eps}
\end{minipage}
\end{tabular}
\end{center}
\caption{Time evolution of the surface roughness and PSD (insets) using \eref{pre_KPZ} with $\omega_k$ given by \eref{omega_compl} (left panel) and \eref{omega_mu} for $\mu=0.75$ (right panel).} \label{fig:W.psd.deBruyn}
\end{figure}

\begin{figure}[t]
\begin{center}
\includegraphics[width=0.6\textwidth]{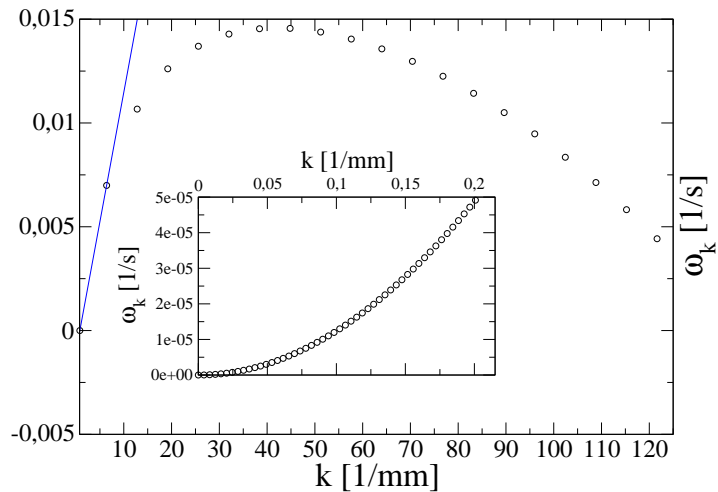}
\end{center}
\caption{Full linear dispersion relation as shown in Fig.\ \ref{bruyn_pre_fig} for $k_D = 5\cdot 10^{-4}$ cm s$^{-1}$. For the meaning of the solid line see the main text. The inset is a zoom showing the $\omega_k \sim k^2$ behaviour for sufficiently small $k$ values.} \label{fig:KS_impos}
\end{figure}

We have performed a numerical integration of an improved version of Eq.\ \eref{pre_KPZ} in which the full linear dispersion relation \eref{omega_compl} is employed, using a stochastic
pseudospectral scheme, see e.g.\ \cite{gallego,nicoli:2008}. Results are shown in Fig.\ \ref{fig:W.psd.deBruyn} for the time evolution of the surface roughness $W(t)$ (root mean square deviation) and the surface structure factor or power spectral density (PSD), $S(k,t)=\langle |\zeta_k(t)|^2\rangle$, of the surface height, $\zeta(x,t)$.
As we see, the roughness increases very fast (close to $W(t) \sim t^{1.8}$) within an intermediate time regime, after which saturation to a stationary value is achieved.
\vskip 0.75cm
As for the time evolution of the PSD curves, they follow a standard Family-Vicsek (FV) behaviour \cite{barabasi}, leading to a global roughness exponent $\alpha = 1.25$, from the asymptotic behaviour $S(k) \sim k^{-(2\alpha+d)}$ for a 1D ($d=1$) interface. The time evolution of the interface height in real space can be appreciated in movie 1 in the supplementary material. As is clear from the images, an initially disordered interface develops mound-like structures that coarsen with time, leading to a long time super-rough morphology. This is inconsistent with the above expectation of KPZ scaling for the present case since, taking into account the values of $V$ and $k_D$, a KS behavior is expected for \eref{omega_compl}.
In the inset of Fig.\ \ref{fig:KS_impos} we show the full dispersion relation \eref{omega_compl} for sufficiently small values of wave-vector at which behaviour is indeed of the KS type. This apparent contradiction can be explained taking into account that such a KS shape (that holds for length scales in the cm range) cannot be observed in practice for the physical length-scales that are accessible in the experiment.
As seen in the figure, the effective behavior of the dispersion relation is not of the KS type but, rather, $\omega_k$ behaves (sub)linearly for the smallest accesible $k$ values. This is made explicit in the main panel of Fig.\ \ref{fig:KS_impos}, where a straight line is shown for reference at the smallest wave-vectors. Thus, for all practical purposes the linear dispersion relation that is experimentally operative does {\em not} behave as $\sim k^2$ but, rather, is closer to a form such as
\begin{equation}
\omega_{\mu}(k) = b_{\mu} |k|^{\mu} - b_2 k^2 ,
\label{omega_mu}
\end{equation}
where $b_{\mu}$ and $b_2$ are positive, effective parameters, and $0 < \mu \leq 1$.

Taking into account the previous paragraphs, we see that \eref{omega_mu} summarizes the main features of the linear dispersion relation for many experimental systems, namely, (i) fast growth at the smallest available $k$ values, $\omega_k \sim |k|^{\mu}$,
(ii) single finite maximum at intermediate $k$ values, and (iii) sufficiently
fast decay at large wave-vectors, at least of the form $\omega_k \sim
-k^2$ or faster (i.e.\ with a larger exponent value). Thus, a simple way to explore the dynamics as predicted by the small slope approximation with an improved dispersion relation consists in considering Eq.\ \eref{pre_KPZ}, but using \eref{omega_mu} rather than
\eref{omega_poly}. We have carried out such type of study for several values
of $\mu$. For the sake of reference, movie 2 in the supplementary material shows the evolution of an interface obeying the KS equation, while movie 3 displays the analogous evolution but as described by Eq.\ \eref{pre_KPZ} using the effective dispersion relation \eref{omega_mu} for $\mu=0.75$. It is apparent that the latter resembles much more closely the dynamics associated with the full dispersion relation \eref{omega_compl}. Quantitative support is provided by the right panel of Fig.\ \ref{fig:W.psd.deBruyn}, where the evolution of the roughness and PSD is shown for Eq.\ \eref{pre_KPZ} with \eref{omega_mu} for $\mu=0.75$. In the inset, we can appreciate the formation of a periodic
surface (signalled as a peak in the PSD) at initial times that grows very rapidly in amplitude, as confirmed by the exponential evolution of the global surface roughness $W(t)$, shown in the main panel of Fig.\ \ref{fig:W.psd.deBruyn},
for times $t\lesssim 15$.
For long enough times, this peak smears out, once the lateral correlation
length is larger than the pattern wavelength. That marks the onset for
nonlinear effects that induce power-law behavior (kinetic roughening) both in
the roughness and in the PSD, approximately as $W(t) \sim t^{1.85}$ and $S(k)
\sim 1/k^{3.5}$, wherefrom we conclude that the associated critical exponents
have values $\beta=1.85$ and $\alpha=1.25$ (thus, $z=\alpha/\beta=0.68$,
not far from $z=\mu=0.75$, an exponent relation which at least holds for
\eref{omega_mu} when $\mu=1$ \cite{nicoli:2008}.) We find it remarkable
that the KPZ nonlinearity is able to stabilize the initial morphological
instability of Eq.\ \eref{pre_KPZ} even in the case of a dispersion relation
such as $\omega_{\mu=0.75}$ that induces such a fast rate of growth. Notice
$z=0.68 < 1$ corresponds effectively to correlations building up at a rate
that is even faster than ballistic. Moreover, the height profile develops
values of the slope at long times that are no longer compatible with the
$|\nabla \zeta(t)|\ll 1$ approximation that led to the derivation of Eq.\ \eref{pre_KPZ} in
the first place. This requires us to reconsider the original moving boundary
problem from a different perspective.

\section{Multivalued interfaces}
\label{sec:pf}

In order to go beyond the small-slope approximation one has to
face the numerical integration of the full moving boundary problem
\eref{n_difusion_eq}-\eref{C_anode}. As mentioned in the introduction,
a method of choice to this end is the so-called phase-field or diffuse
interface approach \cite{gonzalez-cinca:2004}. In this method, one introduces
an auxiliary (phase) field $\phi(\mathbf{r},t)$ that is coupled to the
physical concentration field $c(\mathbf{r},t)$ in such a way as to reproduce
the original moving boundary problem asymptotically, within a perturbation
expansion in the small parameter $\epsilon=W/l_D$. Here, $W$ denotes the
typical size of the region within which the value of $\phi(\mathbf{r},t)$
changes between $\phi=-1$ in the vapor phase and $\phi=+1$ in the solid
phase. The level set $\phi=0$ will thus provide the locus of the aggregate
surface, avoiding the need of explicit front tracking. For our present
physical problem, such an asymptotic equivalence has been recently established
\cite{nicoli:2008b} for the following phase-field reformulation of the
stochastic system \eref{n_difusion_eq}-\eref{C_anode},
\begin{eqnarray}
\tau \partial_t\phi = W^2 \nabla^2 \phi-\partial_\phi \big[f(\phi)-\lambda
u g(\phi)\big] + \sigma_{\phi}, \label{noeqfi1}\\ \partial_t u = \nabla\cdot\left(D
q(\phi)\, \nabla u-\mathbf{j}_{at}\right)-\frac{1}{2}\partial_t
h(\phi) - \nabla \cdot \boldsymbol{\sigma}_{u}, \label{noequ1}
\end{eqnarray}
where, $u=\Omega(c-c^0_{eq})$, $\sigma_{\phi}$ and $\boldsymbol{\sigma}_{u}$ are noises whose variances are related to those of $\mathbf{q}$ and $\chi$ in \eref{n_difusion_eq}-\eref{C_anode} \cite{nicoli:2008b}, and
$f(\phi)$, $g(\phi)$, $q(\phi)$, and $h(\phi)$ are auxiliary functions
that are conveniently chosen (in order to achieve asymptotic convergence)
as $f(\phi)=\phi^4/4-\phi^2/2$, $g(\phi)=\phi-2\phi^3/3+\phi^5/5$,
$q(\phi)= (1-\phi)/2$, and $h(\phi) = \phi$ \cite{Echebarria}. Moreover,
the current $\mathbf{j}_{at}$ generates a so-called anti-trapping
flux that is needed for technical reasons related to convergence
\cite{Echebarria,nicoli:2008b}, and can be taken as $\mathbf{j}_{at}=
8^{-1/2}W(\partial_t\phi)\,\nabla\phi/|\nabla\phi|$. Given model
\eref{noeqfi1}-\eref{noequ1}, one can finally prove \cite{nicoli:2008b}
its convergence to the original moving boundary problem in the thin
interface limit $\epsilon\ll 1$, provided model parameters are related
as \begin{eqnarray} & &d_0 = a_1\frac{W}{\lambda}, \cr & & k_D^{-1} =
a_1\left(\frac{\tau}{\lambda W}-a_2\frac{W}{D}\right), \label{eq_ikd}
\end{eqnarray} where $a_1$ and $a_2$ are constants associated with the
phase-field model, with values
\begin{eqnarray} a_1 & = & 0.8839 , \nonumber \\
a_2 & = & 0.6267 . \label{eq_as}
\end{eqnarray}

There have been previous proposals of phase-field formulations of diffusive \cite{keblinski:1994,keblinski:1994b,keblinski:1996} (and ballistic \cite{keblinski:1995,keblinski:1996}) growth systems of the type of our moving boundary problem. However, in these works the phase-field equations are proposed on a phenomenological basis, rather than being connected with a physical moving boundary problem as in our case. Thus, quantitative comparison is harder to achieve, although many qualitative morphological features can indeed be retrieved.

\begin{figure}
\begin{center}
\begin{tabular}{cc}
\begin{minipage}{0.5\textwidth}
\includegraphics[width=\textwidth]{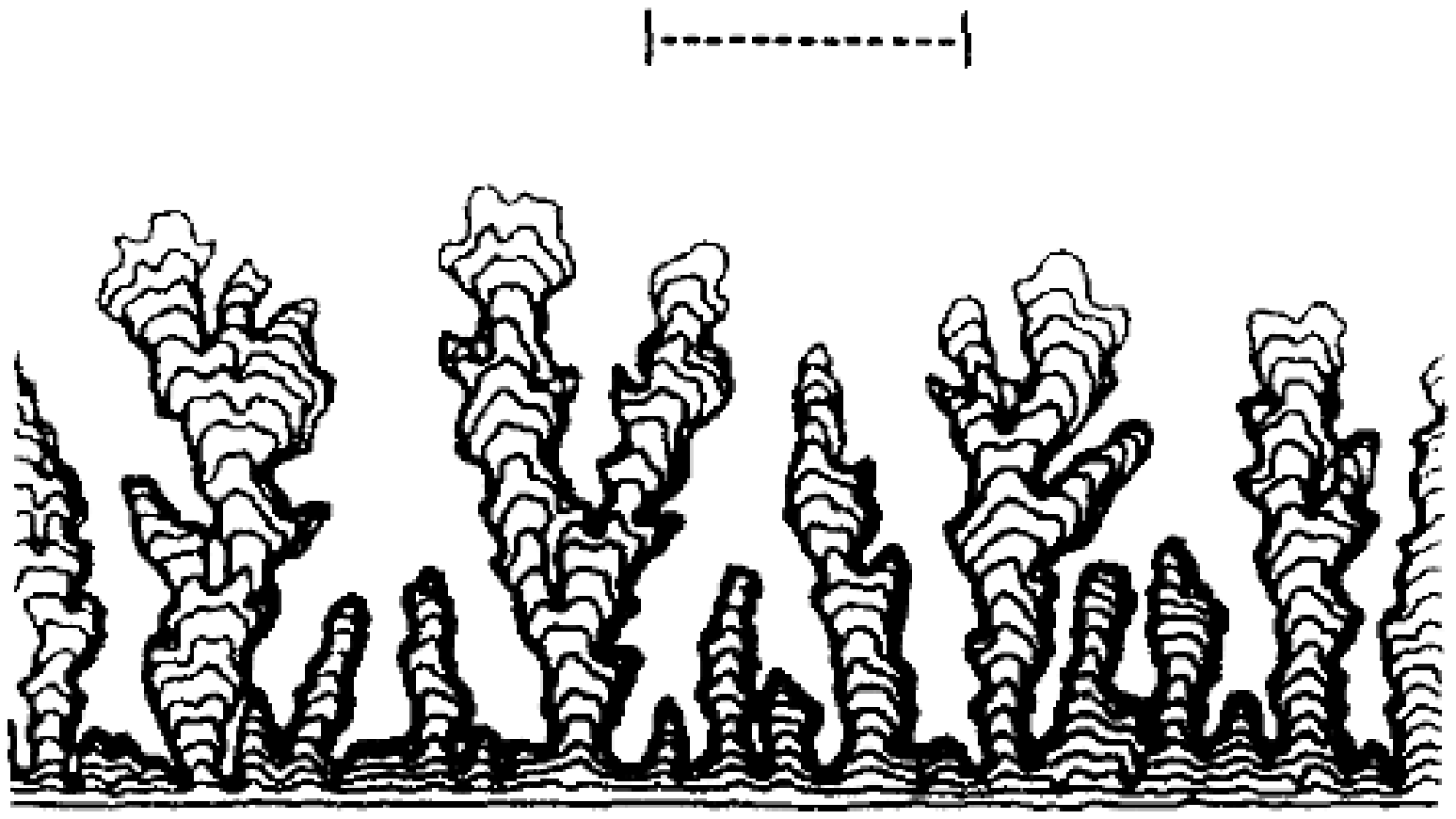}
\end{minipage}
& \begin{minipage}{0.5\textwidth}
\includegraphics[width=\textwidth]{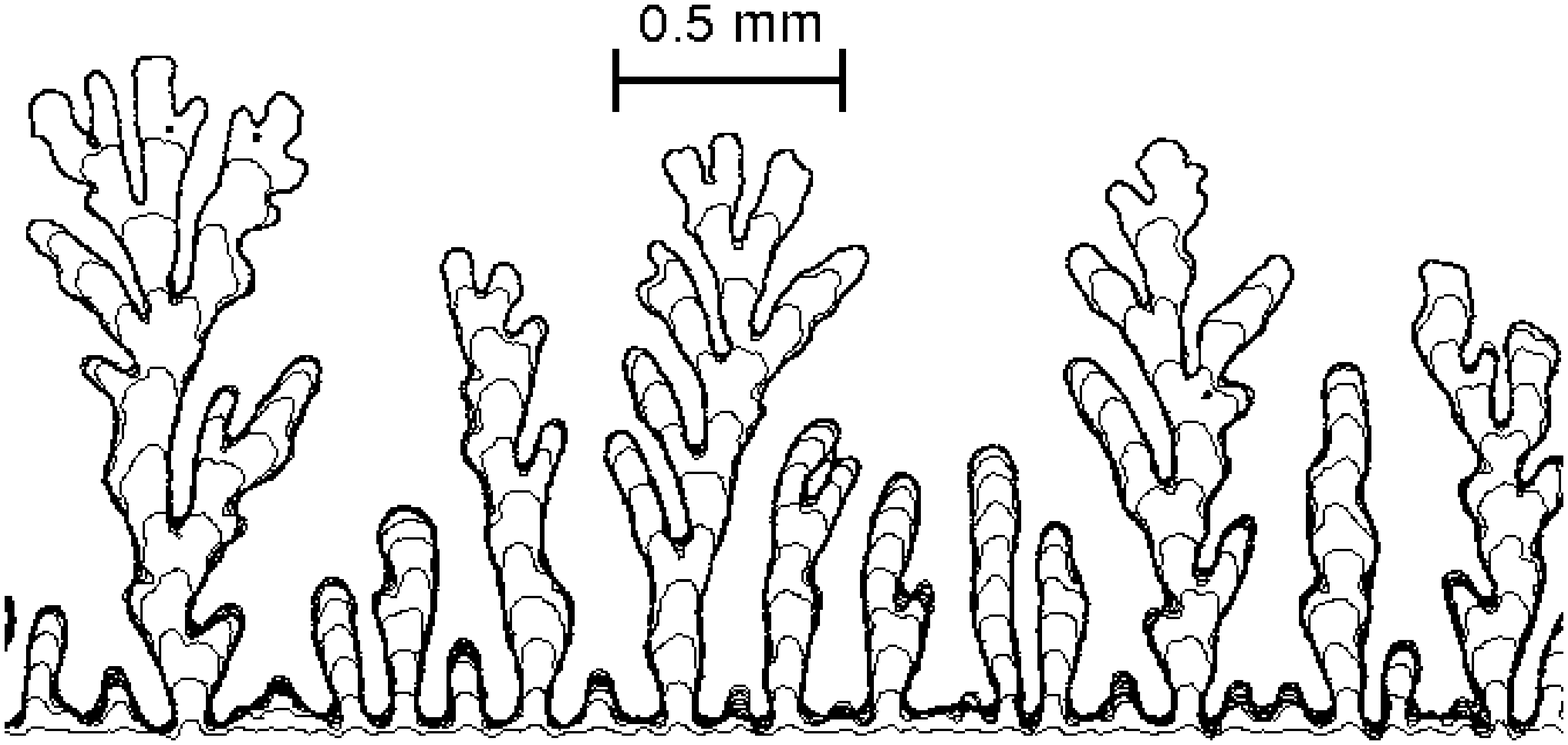}
\end{minipage} \\
\begin{minipage}{0.5\textwidth}
\includegraphics[width=\textwidth]{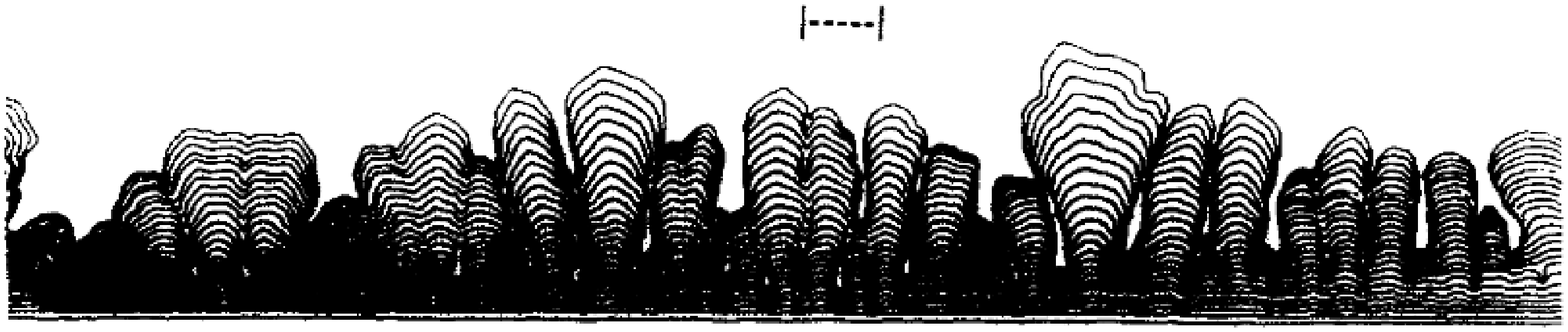}
\end{minipage}
& \begin{minipage}{0.5\textwidth}
\includegraphics[width=\textwidth]{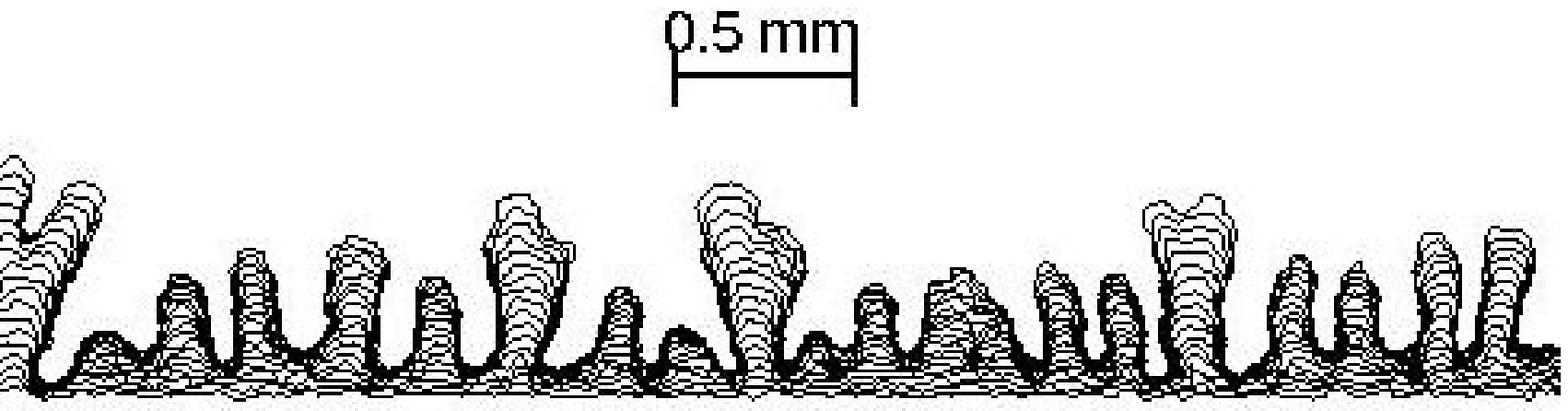}
\end{minipage}
\end{tabular}
\end{center}
\caption{Comparison of morphologies from the experiments on CuSO$_4$ in Ref.\ \cite{kahanda} (left column, taken from the original paper by permission; reference bars represent 0.5 mm) and phase-field simulations using parameters in Table \ref{tablebruyn} (right column). Top row corresponds to fast kinetics ($\eta=-0.50$ V), and bottom row corresponds to intermediate kinetics ($\eta=-0.35$ V).} \label{fig:kahanda}
\end{figure}

We have performed numerical simulations of model \eref{noeqfi1}-\eref{noequ1} using physical parameter values that correspond to the experiments of Cu ECD by Kahanda et al.\ \cite{kahanda} (see parameters in Table \ref{tablebruyn}), for different surface kinetics conditions. Representative morphologies are shown in Fig.\ \ref{fig:kahanda}, where the overpotential value is employed to tune the value of the kinetic coefficient $k_D$.
Notice that in Ref.\ \cite{kahanda} the average growth velocity is overestimated, having been measured in the nonlinear regime, which also overshoots the estimated value of the capillary length. For this reason, the value of the capillary length employed in our simulations (and provided in Table \ref{tablebruyn}) is estimated as smaller than that in \cite{kahanda}. As we can see, for the faster kinetics case (upper row in Fig.\ \ref{fig:kahanda}), indeed the interface becomes multivalued already at early times, leading to a relatively open, branched morphology. As the overpotential increases (thereby decreasing the value of $k_D$, lower row), the morphology is still multivalued, but more compact. We would like to stress the large degree of quantitative agreement between the simulated and experimental morphologies.

\begin{figure}[!h]
\begin{center}
\includegraphics[width=\textwidth]{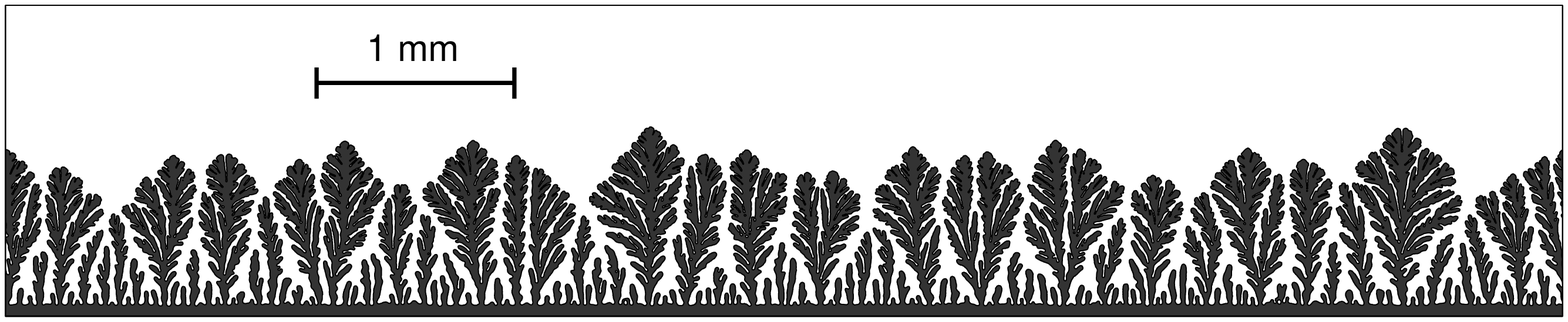}
\includegraphics[width=0.6\textwidth]{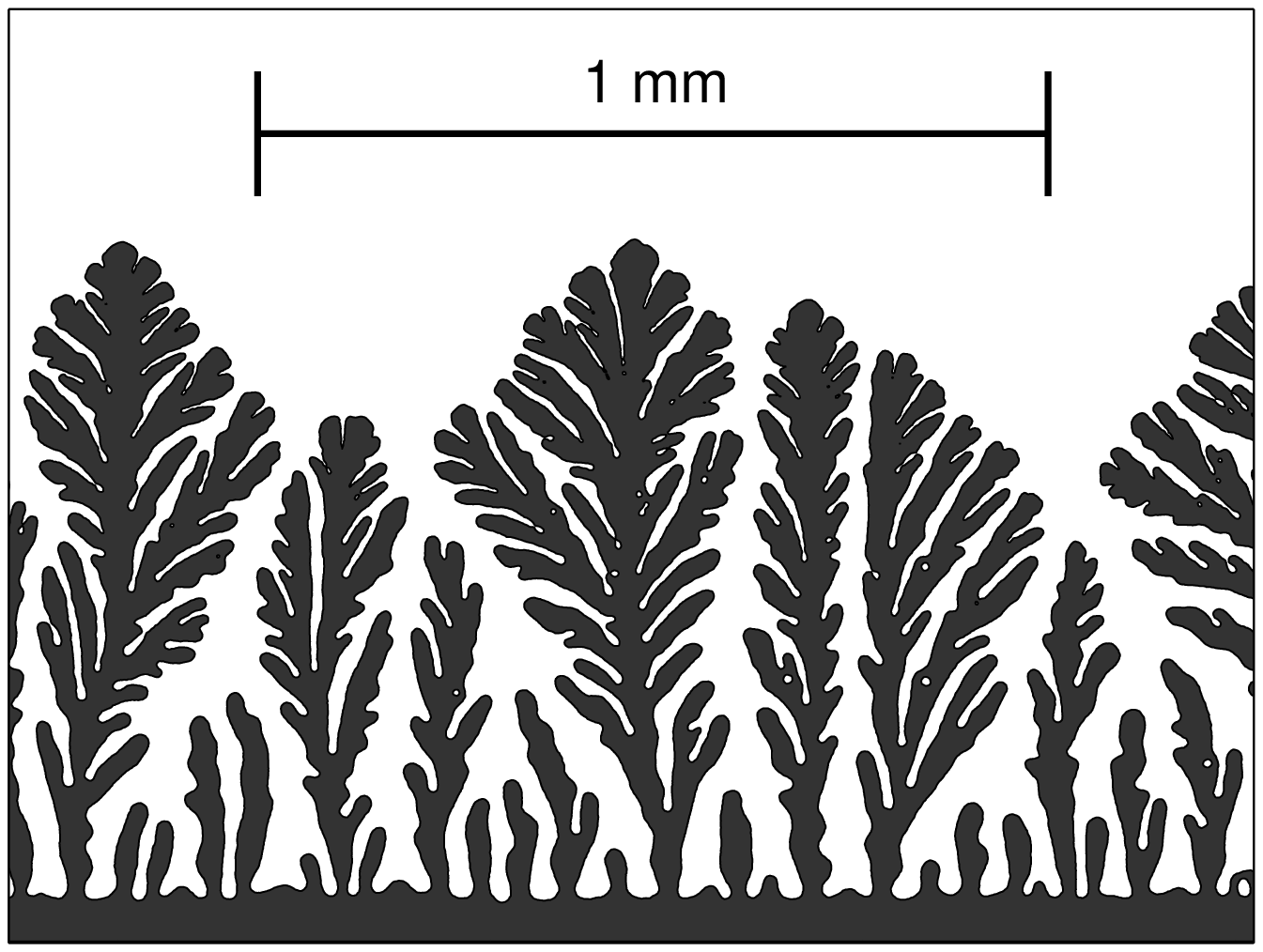}
\end{center}
\caption{Simulated morphology using parameters as in the experiment of Ref.\ \cite{Leger2} on Cu(NO$_3$)$_2$, corresponding to a fast kinetics condition. Lower panel: zoom of a portion of the upper panel.} \label{fig:Leger}
\end{figure}

\begin{figure}[!h]
\begin{center}
\includegraphics[width=0.8\textwidth]{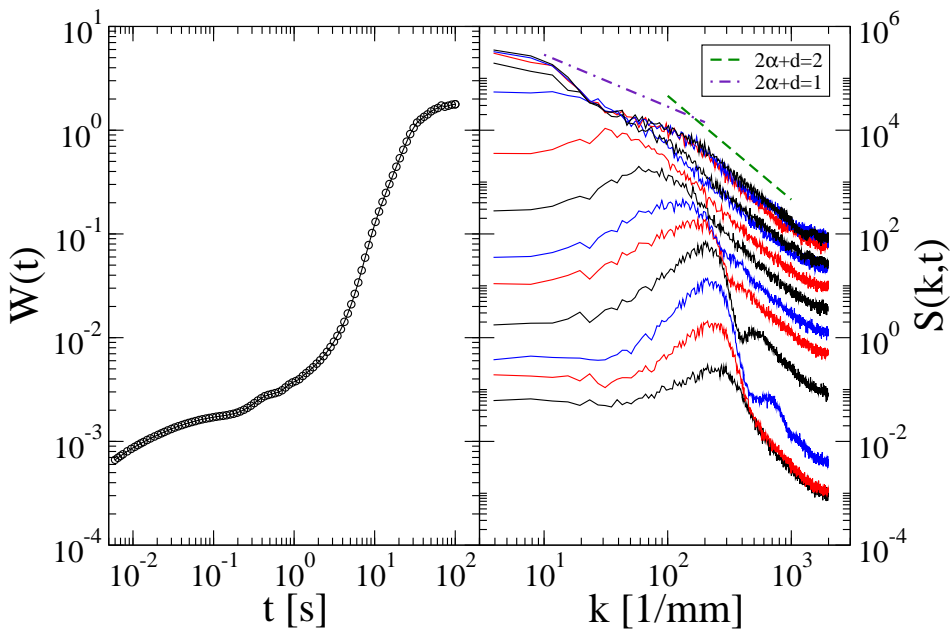}
\end{center}
\caption{Numerical time evolution of the surface roughness (left panel) and PSD (right panel) for the same parameters as in Fig.\ \ref{fig:Leger}. The PSD curves are for increasing times from bottom to top. Notice the time shift of the PSD curves for high values of $k$ (small length scales) signalling anomalous scaling, see the text for details. For reference, the dashed line has slope $-2$ while the dash-dotted line has slope $-1$.} \label{fig:Leger_W_PSD}
\end{figure}

Our phase-field model (and the corresponding moving boundary problem that it represents) has moreover a large degree of universality, in the sense that it can also reproduce quantitatively other ECD systems. An example is provided by Fig.\ \ref{fig:Leger}, which shows simulations for ECD growth from a Cu(NO$_3$)$_2$ solution, as in Ref.\ \cite{Leger2}, see experimental parameters in Table \ref{tablebruyn}. This dense branching morphology corresponds to a fast kinetics condition for which the single valued approximation breaks down already for short times. Fig.\ \ref{fig:Leger_W_PSD} collects the time evolution of the surface roughness and of the PSD function for the same parameters as in Fig.\ \ref{fig:Leger}.
The roughness grows rapidly within a transient regime, that is later followed by stabilization and rapid saturation to a stationary value. This result is similar to those obtained within the small slope approximation. However, the behavior of the PSD differs dramatically. Thus, for very small times the linear instability shows up in the finite maximum of the curves. For intermediate times the PSD's shift upwards with time, a behaviour that is associated with anomalous kinetic roughening properties \cite{castro1998asn,SchwarLett1} that contrasts with the standard FV scaling found above in which the PSD values are time independent for length scales that are below the correlation length. Note that at these intermediate times the single valued approximation does not describe the features of the interface at all. Actually, the single valued approximation to the height field (not shown) presents abundant large jumps that are possibly responsible for the effective anomalous scaling \cite{asikainen2002}. However, for sufficiently long times branches grow laterally closing the interbranch spacing, and the validity of the single valued approximation is restored. Thus, the PSD curves for the longest times again display Family-Vicsek behavior for the largest $k$ values. A crossover between $\alpha=0.5$ for small scales and $\alpha=0$ (log) for intermediate scales can be read off from the figure.

Note that, for the longest distances in our system, scale invariance seems to break down, since no clear power law can be distinguished for, say, $k < 30$ mm$^{-1} \simeq l_D^{-1}$.
In order to properly characterize the behavior of the interface at these scales, simulations are required for system sizes that are much larger the diffusion length. We hope to report on this in the future.

\section{Discussion and Conclusions}

In this paper we have studied the physics of non-conserved growth from a diffusive source. We have focused on, both, morphological features and dynamical properties, making special emphasis in the emergence of a long time structure arising from a well characterized linearly unstable regime. Thus, we have dealt with the problem from two different but complementary points of view: the linearized dispersion relation obtained from the moving boundary description of the problem (as presented in Ref.\ \cite{nicoli:2008}) and the
use of a phase-field equivalent description of the full original physical system.

This complementary study has allowed us to understand the evolution of the problems in two well separated regimes: one at short times, well characterized by the linear dispersion relation, and another one at long times in which steep slopes may eventually develop and break down the simplistic description through a single valued height field. We want to emphasize that the comparison has been done quantitatively. Thus, in Sec.\ II we have shown how our theoretical predictions allow to explain some experimental results reported in \cite{bruynpre}. At variance with the data analysis performed in this reference, our comparison is not based on a multidimensional fit of the parameters but arises, rather, from a prediction for the form of the dispersion relation compatible with the available experimental data.

In order to develop insight into the dynamics decribed by the complex full dispersion relation \eref{omega_compl}, we have proposed a simple description of the evolving interface in terms of an effective dispersion relation, in which the exponent $\mu$ of the destabilizing term comprises all the relevant information on the behavior of the morphological instability at the largest length scales accesible in the system.
Hence, the simple description of the data given by Eq.\ \eref{pre_KPZ} using \eref{omega_compl} provides a promising testbed for future theoretical analysis of many problems sharing similar ingredients with ours. Another benefit of this characterization of the linear regime is to provide an explanation for the scarcity of experimental evidence of the celebrated KPZ scaling (or even its generalization for unstable systems given by the KS equation). As we have demonstrated, the difficulty of actually observing the experimental systems in the KPZ universality class is due to the fact that the physical conditions for the observation of that behavior cannot be fulfilled for realistic (experimentally motivated) values of the parameters, as they typically require system sizes that are beyond experimental feasibility.

Besides, with the help of the phase-field model we have also performed a systematic analysis of the experiments reported in \cite{kahanda}. Our comparison not only reproduces qualitatively well the morphologies obtained in those experiments but also predicts the existence of a scaling regime characterized by anomalous scaling. This anomalous scaling deserves further analysis but, preliminarily, can be understood in terms of the emergence of large slopes for times following appearance of overhangs and multivalued interfaces. As we have shown, our model sheds some light on the characterization of the experiments reported in Refs.\ \cite{kahanda,Leger2} and the role of the different mechanisms involved. The quantitative estimation of the parameters appearing in our theoretical framework leads also to remarkable agreement with the morphological structure of the observed
 deposits.

Finally, we want to stress that, although the formulation of the electrochemical deposition process with a phase-field theory has been addressed before in the literature \cite{keblinski:1994,keblinski:1994b,keblinski:1996,keblinski:1995}, our model is not constructed phenomenologically, but is, rather, an equivalent formulation of the physical equations (diffusion, electrochemical reduction at the cathode, etc.) that reduces to them in the thin-interface limit. In addition, our formulation allows us to do a quantitative comparison with experiments which is far from the capabilities of the mentioned theories. Notwithstanding, the promising results presented above still need more detailed comparison and analysis (through more extensive numerical experiments), which will be the aim of future work. This will focus particularly on the appearance of the anomalous scaling regime and, from the morphological point of view, on the influence, cooperation and/or competition of different physical mechanisms that are naturally included in our theory.

\section{Acknowledgments}

This work has been partially supported by MEC (Spain), through Grants No.\ FIS2006-12253-C06-01 and No.\ FIS2006-12253-C06-06, and by CAM (Spain) through Grant No.\ S-0505/ESP-0158. M.\ N.\ acknowledges support by the FPU programme (MEC) and by Fundaci\'on Carlos III.

\section*{References}

\bibliography{nicoli}

\begin{thebibliography}{10}

\bibitem{cvd}
K. Jensen and W. Kern,  in {\em Thin Film Processes II}, edited by J. Vossen
  and W. Kern (Academic, Boston, 1991).

\bibitem{ecd}
J. Bockris and A. Reddy, {\em Modern electrochemistry} (Plenum/Rosetta, New
  York, 1970).

\bibitem{mbe}
A. Pimpinelli and J. Villain, {\em {Physics of Crystal Growth}} (Cambridge
  University Press, Cambridge, 1998).

\bibitem{matsushita2005cfb}
M. Matsushita {\it et~al.}, Biofilms {\bf 1},  305  (2005).

\bibitem{ball}
P. Ball, {\em {The self-made tapestry: pattern formation in nature}} (Oxford
  University Press, Oxford, 1999).

\bibitem{barabasi}
A.-L. Barab\'asi and H.~E. Stanley, {\em Fractal concepts in surface growth}
  (Cambridge University Press, Cambridge, 1995).

\bibitem{cuerno:2007}
R. Cuerno {\it et~al.}, Eur. Phys. J. Special Topics {\bf 146},  427  (2007).

\bibitem{castro1998asn}
M. Castro, R. Cuerno, A. S{\'a}nchez, and F. Dom{\'\i}nguez-Adame, Phys. Rev. E
  {\bf 57},  2491  (1998).

\bibitem{castro2000mbd}
M. Castro, R. Cuerno, A. S{\'a}nchez, and F. Dom{\'\i}nguez-Adame, Phys. Rev. E
  {\bf 62},  161  (2000).

\bibitem{nicoli:2008}
M. Nicoli, M. Castro, and R. Cuerno, Phys. Rev. E {\bf 78},  21601  (2008).

\bibitem{gonzalez-cinca:2004}
R. Gonz\'alez-Cinca {\it et~al.},  in {\em Advances in Condensed Matter and
  Statistical Physics}, edited by E. Korutcheva and R. Cuerno (Nova Science
  Publishers, New York, 2004).

\bibitem{plapp2000mrw}
M. Plapp and A. Karma, Phys. Rev. Lett. {\bf 84},  1740  (2000).

\bibitem{karma1996pfm}
A. Karma and W. Rappel, Phys. Rev. E {\bf 53},  3017  (1996).

\bibitem{wang1993tcp}
S. Wang {\it et~al.}, Physica D {\bf 69},  189  (1993).

\bibitem{elder2002mec}
K. Elder, M. Katakowski, M. Haataja, and M. Grant, Phys. Rev. Lett. {\bf 88},
  245701  (2002).

\bibitem{granasy2003gdd}
L. Granasy {\it et~al.}, Nature Mater. {\bf 2},  92  (2003).

\bibitem{karma1998ssg}
A. Karma and M. Plapp, Phys. Rev. Lett. {\bf 81},  4444  (1998).

\bibitem{castro2003pfa}
M. Castro, Phys. Rev. B {\bf 67},  35412  (2003).

\bibitem{dube2000cda}
M. Dub{\'e} {\it et~al.}, Eur. Phys. J. B {\bf 15},  701  (2000).

\bibitem{alanissila2004pfm}
T. Ala-Nissila, S. Majaniemi, and K. Elder, Lecture Notes in Physics  357
  (2004).

\bibitem{campelo2007mcd}
F. Campelo and A. Hern{\'a}ndez-Machado, Phys. Rev. Lett. {\bf 99},  88101
  (2007).

\bibitem{Jansen}
C.~H. J.~V. den Breckel and A.~K. Jansen, J. Cryst. Growth {\bf 43},  364
  (1978).

\bibitem{Palmer88}
B.~J. Palmer and R.~G. Gordon, Thin Solid Films {\bf 158},  313  (1988).

\bibitem{Bales}
G.~S. Bales, A.~C. Redfield, and A. Zangwill, Phys. Rev. Lett. {\bf 62},  776
  (1989).

\bibitem{Bard}
A.~J. Bard and L.~R. Faulkner, {\em Electrochemical methods} (John Wiley and
  Sons, New York, 1980).

\bibitem{Kashchiev}
D. Kashchiev and A. Milchev, Thin Solid Films {\bf 28},  189  (1975).

\bibitem{mullins:1957}
W.~W. Mullins, J. Appl. Phys. {\bf 28},  333  (1957).

\bibitem{mullins:1959}
W.~W. Mullins, J. Appl. Phys. {\bf 30},  77  (1959).

\bibitem{ourprl}
R. Cuerno and M. Castro, Phys. Rev. Lett. {\bf 87},  236103  (2001).

\bibitem{cuerno:2002}
R. Cuerno and M. Castro, Physica A {\bf 314},  192  (2002).

\bibitem{Schilardi}
P.~L. Schilardi, O. Azzaroni, R.~C. Salvarezza, and A.~J. Arvia, Phys. Rev. B
  {\bf 59},  4638  (1999).

\bibitem{cuerno:2004}
R. Cuerno and L. V\'azquez,  in {\em Advanced in Condensed Matter and
  Statistical Physics}, edited by E. Korutcheva and R. Cuerno (Nova Science
  Publishers, New York, 2004).

\bibitem{Ojeda}
F. Ojeda, R. Cuerno, R. Salvarezza, and L. V\'azquez, Phys. Rev. Lett. {\bf
  84},  3125  (2000).

\bibitem{ojeda:2003}
F. Ojeda {\it et~al.}, Phys. Rev. B {\bf 67},  245416  (2003).

\bibitem{nicoli:2008b}
M. Nicoli, M. Castro, M. Plapp, and R. Cuerno, J. Stat. Mech.  submitted
  (2008).

\bibitem{bruynpa}
G. White and J. de~Bruyn, Physica A {\bf 239},    (1997).

\bibitem{bruynpre}
J. de~Bruyn, Phys. Rev. E {\bf 53},  5561  (1996).

\bibitem{kahanda}
G.~L. M. K.~S. Kahanda, X.-q. Zou, R. Farrell, and P.-z. Wong, Phys. Rev. Lett.
  {\bf 68},  3741  (1992).

\bibitem{Leger2}
C. L\'eger, J. Elezgaray, and F. Argoul, Phys. Rev. E {\bf 58},  7700  (1998).

\bibitem{gallego}
R. Gallego, M. Castro, and J.~M. L\'{o}pez, Phys. Rev. E {\bf 76},  051121
  (2007).

\bibitem{Echebarria}
B. Echebarria, R. Folch, A. Karma, and M. Plapp, Phys. Rev. E {\bf 70},  061604
   (2004).

\bibitem{keblinski:1994}
P. Keblinski {\it et~al.}, Phys. Rev. E {\bf 49},  R937  (1994).

\bibitem{keblinski:1994b}
P. Keblinski, A. Maritan, F. Toigo, and J.~R. Banavar, Phys. Rev. E {\bf 49},
  R4795  (1994).

\bibitem{keblinski:1996}
P. Keblinski {\it et~al.}, Phys. Rev. E {\bf 53},  759  (1996).

\bibitem{keblinski:1995}
P. Keblinski, A. Maritan, F. Toigo, and J.~R. Banavar, Phys. Rev. Lett. {\bf
  74},  1783  (1995).

\bibitem{SchwarLett1}
S. Huo and W. Schwarzacher, Phys. Rev. Lett. {\bf 86},  256  (2001).

\bibitem{asikainen2002}
J. Asikainen {\it et~al.}, Eur. Phys. J. B {\bf 30},  253  (2002).

\end{thebibliography}
\end{document}